\documentclass[amsmath,amssymb,twocolumn]{revtex4}

\usepackage{epsf}

\begin{document}

\title{Collinear helium under periodic driving: stabilization of the
  asymmetric stretch orbit}

\author{Peter Schlagheck$^1$}
\author{Detlef Pingel$^2$}
\author{Peter Schmelcher$^{2,3}$}
\affiliation{$^1$Institut f{\"u}r Theoretische Physik, Universit{\"a}t Regensburg,
  93040 Regensburg, Germany}
\affiliation{$^2$Theoretische Chemie, Im Neuenheimer Feld 229,
  Universit{\"a}t Heidelberg, 69120 Heidelberg, Germany}
\affiliation{$^3$ Physikalisches Institut, Philosophenweg 12, Universit{\"a}t
  Heidelberg, 69120 Heidelberg, Germany}

\date{\today}

\begin{abstract}

The collinear $eZe$ configuration of helium, with the electrons on opposite
sides of the nucleus, is studied in the presence of an external
electromagnetic (laser or microwave) field.
We show that the classically unstable ``asymmetric stretch'' orbit, on which
doubly excited intrashell states of helium with maximum interelectronic angle
are anchored, can be stabilized by means of a resonant driving where the
frequency of the electromagnetic field equals the frequency of Kepler-like
oscillations along the orbit.
A static magnetic field, oriented parallel to the oscillating electric field
of the driving, can be used to enforce the stability of the configuration with
respect to deviations from collinearity.
Quantum Floquet calculations within a collinear model of the driven
two-electron atom reveal the existence of nondispersive wave packets localized
on the stabilized asymmetric stretch orbit, for double excitations
corresponding to principal quantum numbers of the order of 
$N {\scriptstyle {> \atop ^\sim}} 10$.

\end{abstract}


\maketitle

\section{Introduction}

The correlated dynamics of two-electron atoms under external electromagnetic
driving represents a fascinating topic of atomic physics.
The outstanding example in this context is double ionization of helium in
the presence of a strong laser pulse.
At not too high intensities of the laser (near $10^{15}$ W$/$cm$^2$
\cite{FitO92PRL,WalO94PRL}), the electrons are, contrary to natural 
expectation, {\em not} emitted in a sequential process -- which was clearly
manifest in the original experiments \cite{FitO92PRL,WalO94PRL}.
Further experimental investigations (e.g.,
\cite{LarTalChi98JPB,WebO00PRL,FeuO01PRL}) and a number of theoretical studies
(e.g., \cite{Cor93PRL,BecFai96JPB,WatO97PRL,BecFai00PRL,YudIva01PRA}) have
revealed that a nontrivial multistep process is responsible for nonsequential
double ionization in this intensity regime, involving tunnel ionization of one
of the electrons, recollision of this outer electron with the ionic core,
followed by subsequent excitation and ionization of the inner electron.

In the regime of highly excited states, complex dynamics can also take place
in the presence of {\em moderate} electromagnetic fields, which are not strong
enough to fully ionize the atom, but substantially perturb the electronic
motion.
A striking example in this context is the occurence of nondispersive wave
packets in microwave-driven hydrogen:
Rydberg wave packets, which are normally subject to spreading and collapse
after a limited number of Kepler cycles, can be stabilized and ``kept in
shape'' over a practically arbitrary amount of time by means of a resonant
electrical driving (with linear \cite{Buc93,BucDel95PRL} or circular
polarization \cite{BiaKalEbe94PRL,ZakDelBuc95PRL,KalEbe96PRL}). 
This field needs to be phase-matched with the wave packet in such a way that
components associated with deviating classical trajectories are driven back to
the resonant Kepler orbit.
For a wave packet that performs parabolic-state-like oscillations along the
axis of a linearly polarized field for instance \cite{Buc93,BucDel95PRL}, this
means that the external force on the electron needs to be directed towards the
nucleus at the outer turning point of the classical motion, and away from it
at the inner turning point (i.e., when the electron collides with the nucleus).
The wave packet then corresponds to a Floquet eigenstate of the periodically
driven system, which is localized in phase space on the regular island that is
associated with the above nonlinear resonance.
The coupling to the ``chaotic sea'' outside the island, and subsequently also
to the ionization continuum, is mediated via a classically forbidden tunneling
process through the dynamical phase space barriers of the island, what leads
to almost ``eternal'' lifetimes of such wave packets 
\cite{Buc93,BucDel95PRL,ZakDelBuc95PRL}.

The concept of nondispersive wave packets, which is of potential interest in
the context of quantum control, can be generalized to the correlated dynamics
in two-electron atoms.
This was indeed shown for the classically stable ``frozen planet''
configuration of helium \cite{RicWin90PRL,RicO92JPB}, where both electrons are
located on the same side of the nucleus with asymmetric excitation.
The application of a linearly polarized microwave perturbation to this
configuration induces analogous islands of regular motion in phase space,
which arise here from a resonant driving of the outer electron
\cite{SchBuc98JPB,SchBuc99PhD}.
A static electric field, however, is needed to ensure classical stability
with respect to deviations from collinearity.
The existence of nondispersive two-electron wave packets localized on these
resonance islands has indeed been revealed in Floquet calculations within a
quantum model that represents the analog of the collinear two-electron
configuration \cite{SchBuc99EPL,SchBuc03EPD}.

In this contribution, we examine to which extent the mechanism that leads to
nondispersive wave packets can be used to {\em stabilize} a doubly
excited configuration -- i.e., to enhance the lifetime of the associated
autoionizing states by increasing the amplitude of the external
electromagnetic field.
Originally, the phenomenon of stabilization was discussed in the context of
one-electron atoms in ultra-intense high-frequency laser fields (``adiabatic
stabilization''), where it was shown that the ionization rate of the atom
decreases with increasing laser intensity \cite{PonGav90PRL} (without,
however, taking into account relativistic effects; see \cite{KylO00PRL}).
The effect that we are aiming at is more of {\em intermediate} nature
(i.e., it manifests itself at moderate intensities of the electromagnetic
field)
%(in a similar way as ``dynamical stabilization'' [\ldots]) 
and relies on the structure of the underlying classical phase space.
The basic question is to which extent the driving field can create dynamical
phase space barriers (invariant tori) around a periodic orbit that would be
unstable without the external perturbation.
Floquet eigenstates that are locally quantized on this stabilized orbit are
then semiclassically protected against decay, which enhances their lifetime
compared to the unperturbed atom.

A natural candidate for this stabilization mechanism is the ``asymmetric
stretch'' orbit \cite{EzrO91JPBL}.
In this orbit, the electrons are located on opposite sides of the nucleus and
perform slightly perturbed Kepler oscillations with opposite phase; whenever
one of the electrons hits the nucleus, the other one reaches the outer turning
point of the orbit.
The dynamics along this orbit is stable with respect to transverse bending
perturbations, but unstable against deviations in axial direction.
As was shown by Richter and Wintgen \cite{RicWin93JPB}, intrashell resonances
with maximum interelectronic angle are predominantly scarred along this orbit
-- which can be seen as a consequence of the prominent role that this orbit
plays in the semiclassical quantization of helium \cite{EzrO91JPBL}.

Due to the phase shift between the Kepler-like oscillations, the stabilization
mechanism that was used to create nondispersive wave packets in driven
hydrogen can now be applied {\em simultaneously to both electrons}:
We expose the collinear configuration to a coaxially polarized electromagnetic
field the force of which is directed towards the nucleus at the outer turning
point of each electron, and outwards when the electron collides with the
nucleus.
Indeed, we shall see in Section \ref{CL} that the asymmetric stretch orbit can
thereby be stabilized within the phase space of collinear motion, if the
amplitude of the driving field is appropriately chosen.
In the presence of the time-periodic driving, the orbit is no longer stable
with respect to deviations from collinearity -- due to mixing of states with
different angular momentum.
We shall argue in Section \ref{2D}, however, that a static magnetic field
can be used to enforce the transverse stability of the driven configuration.
In Section \ref{QM} finally, the basic properties of nondispersive
two-electron wave packets localized on the stabilized asymmetric stretch orbit
are elaborated using a one-dimensional model that represents the quantum
analog of the collinear $eZe$ configuration.
We shall point out that these wave packet states evolve diabatically from the
unperturbed asymmetric stretch state, and that their lifetimes exhibit a local
maximum at finite values of the field amplitude.

\section{Classical stabilization of the collinear asymmetric stretch orbit}

\label{CL}

In atomic units, which are used throughout this paper, the classical
Hamiltonian of the electromagnetically driven helium atom reads
\begin{eqnarray}
H & = & \frac{{\bf p}_1^2}{2} \; + \; \frac{{\bf p}_2^2}{2} 
  \; - \; \frac{Z}{|{\bf r}_1|} \; - \; \frac{Z}{|{\bf r}_2|}
  \; + \; \frac{1}{|{\bf r}_1 - {\bf r}_2|} \nonumber \\
 && \; + \; (z_1 + z_2) F \cos (\omega t + \varphi). \label{eq:hcl}
\end{eqnarray}
Here, ${\bf r}_i = (x_i, y_i, z_i)$ and ${\bf p}_i = (p_{xi}, p_{yi}, p_{zi})$
denote the position and momentum of electron $i = 1,2$, respectively, $Z = 2$
is the nuclear charge, and $F$ and $\omega$ represent the amplitude and frequency
of the external driving field which is linearly polarized along the $z$ axis.
In analogy to driven hydrogen \cite{LeoPer78PRL} as well as to the
unperturbed helium atom \cite{Per77ACP}, the Hamiltonian (\ref{eq:hcl})
exhibits general scaling laws: 
The classical dynamics generated by (\ref{eq:hcl}) remains invariant if all
variables and parameters of the system are transformed according to
\begin{subequations}
\label{eq:sc}
\begin{eqnarray}
{\bf r}_i & \longmapsto & \nu^2 \, {\bf r}_i \quad (i = 1,2), \\
{\bf p}_i & \longmapsto & \nu^{-1} \, {\bf p}_i \quad (i = 1,2), \\
%\label{scp} \\
t & \longmapsto & \nu^3 \, t, \\
{\bf F} & \longmapsto & \nu^{-4} \, F, \\
\omega & \longmapsto & \nu^{-3} \, \omega, \\
H & \longmapsto & \nu^{-2} \, H, 
\end{eqnarray}
\end{subequations}
where $\nu$ represents an arbitrary, real positive quantity.
We shall therefore restrict the classical analysis to a fixed value $\omega = 1$ of
the driving frequency, and use the above scale invariance (\ref{eq:sc}) to
deduce the actual dynamics at the energy range of interest.
$F_0$ denotes, in the following, the value of the ``scaled'' field amplitude
$F$ at driving frequency $\omega = 1$.

In the following, we shall concentrate on the invariant subspace of collinear
motion along the $z$ axis, with the electrons on {\em opposite} sides of the
nucleus.
Such a collinear $eZe$ motion is generated by initial conditions of the form
$x_i(t=0) = y_i(t=0) = p_{xi}(t=0) = p_{yi}(t=0) = 0$ for $i=1,2$, and
$z_1(t=0) > 0$ whereas $z_2(t=0) < 0$.
%\begin{subequations}
%\begin{equation}
%x_i(t=0) = y_i(t=0) = p_{xi}(t=0) = p_{yi}(t=0) = 0
%\end{equation}
%for $i = 1,2$, and 
%\begin{equation}
%z_1(t=0) > 0, z_2(t=0) < 0.
%\end{equation}
%\end{subequations}
Due to the Coulomb singularity in the interaction, the electrons cannot pass
the nucleus at the origin.
This implies that $z_1(t) > 0$ and $z_2(t) < 0$ for all times $t>0$.

To ensure a stable numerical integration of the classical equations of motion,
we perform a Kustaanheimo-Stiefel transformation \cite{KusSti65JRAM}.
This procedure, which is described in Appendix \ref{ap:ks} for the special
case of collinear motion, introduces new phase space variables which do not
diverge at electron-nucleus collisions.
Triple collisions -- i.e., the simultaneous encounter of both electrons at the
nucleus -- cannot be regularized in such a way;
they represent ``true'' singularities of the dynamics, where different
manifolds of trajectories merge together without a well-defined continuation
\cite{Sie41AM}.

\begin{figure}
\begin{center}
\leavevmode
\epsfxsize8cm
\epsfbox{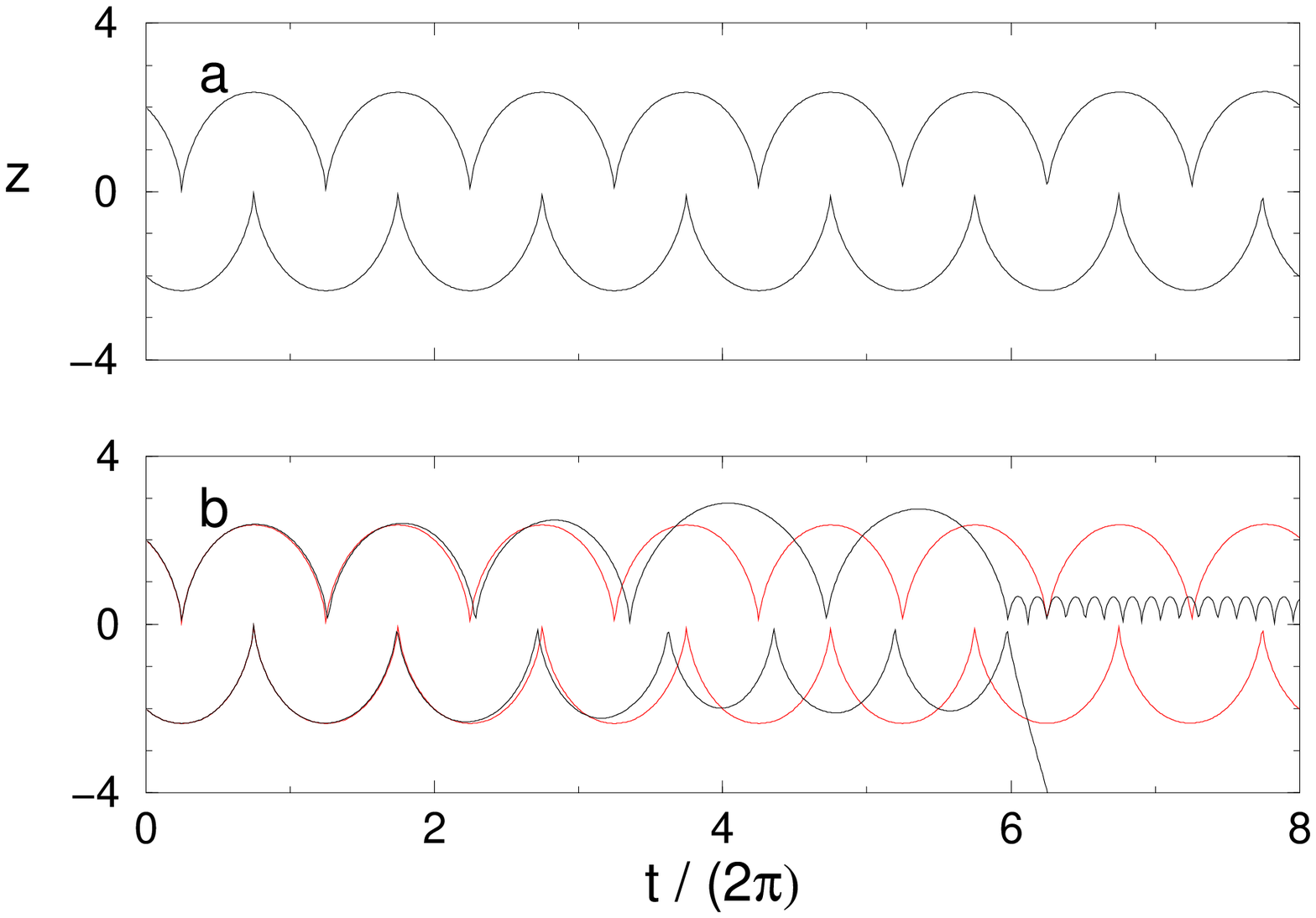}
% page: 600 x 400
\end{center}
\caption{a) The asymmetric stretch orbit of the collinear $eZe$ configuration.
  Plotted are the coordinates of the two electrons ($z_1 > 0$ and $z_2 < 0$)
  as a function of time, with the nucleus resting at $z = 0$.
  The scaling (see Eq.~(\ref{eq:sc})) is chosen such that the frequency of the
  orbit equals $\omega = 1$.\\
  b) A small change of the initial condition (the ``upper'' electron is
  displaced by $\Delta z_1 = 0.01$) leads to an appreciable deviation from the
  asymmetric stretch orbit (shown in grey in the background) within a couple
  of oscillation cycles.
  At $t / (2 \pi) \simeq 6$, the configuration ionizes after a nearby triple
  collision.
\label{fg:as}}
\end{figure}

Correspondingly, the Kolmogorov-Arnold-Moser (KAM) theorem cannot be applied
to this system, which is reflected by the fact that the classical dynamics of
the unperturbed three-body problem (i.e., in the field-free case $F_0=0$) is
fully chaotic, even in the limit of very weak electron-electron interaction
(which would correspond here to very small $1/Z$)
\cite{EzrO91JPBL,BaiGuYua98PsD}.
Indeed, it is possible to represent each periodic orbit by a unique sequence
of symbols, which is basically determined by the order in which the electrons
hit the nucleus \cite{EzrO91JPBL}.
The simplest orbit with respect to this symbolic code is the ``asymmetric
stretch'' orbit, shown in Fig.~\ref{fg:as}(a), in which the electrons collide
with the nucleus in a perfectly alternating way.
Since in this configuration the electrons avoid a simultaneous encounter at
the nucleus in the best possible way, the stability exponent of this orbit is
rather low compared to other, more complicated periodic orbits
\cite{EzrO91JPBL}.

Nevertheless, a small deviation from the orbit's initial condition leads to 
disintegration of the configuration on rather short time scales.
This is illustrated in Fig.~\ref{fg:as}(b), which shows what happens when one
of the electrons is displaced in configuration space by an amount of $\Delta z =
0.01$:
After a few cycles, the Kepler-like oscillations of the electrons get out of
phase, which leads to a nearby triple collision where a large amount of energy
is transferred between the electrons, and to subsequent (single) ionization of
the atom.

The external time-dependent electric field is now applied to this
configuration in order to compensate this destabilization phenomenon.
As in the case of nondispersive wave packets in driven hydrogen, the relative
phase between the field and the electronic motion is adjusted such that the
field forces the electron towards the nucleus at the outer turning point of
the Kepler-like oscillation, and away from it at the inner turning point.
In this way, deviations from the periodic orbit are counterbalanced by the
field.
Since their oscillations take place with opposite phase, the phase-matching
condition can be {\em simultaneously} satisfied for both electrons.
For the particular type of initial condition in Fig.~\ref{fg:as}, where
$z_1(0) = - z_2(0)$ and $p_{z1}(0) = p_{z2}(0) < 0$, we need to choose $\varphi = \pi/2$
as initial field phase in Eq.~(\ref{eq:hcl}).

Due to the high dimensionality of the extended phase space, which is spanned
by the positions and momenta of the electrons and by the phase of the external
driving, regular islands that result from this stabilization mechanism cannot
be visualized by means of Poincar{\'e} surfaces of section.
To identify the periodic orbit at finite field amplitude $F_0 \neq 0$, we employ an
iterative procedure which is described in Appendix \ref{ap:po}.
The key ingredient to this procedure is the fact that initial conditions of
asymmetric stretch orbits at different field amplitudes $F_0$ and 
$F_0 + \delta F_0$ lie close to each other in phase space -- i.e., within each
other's linear neighbourhood -- as long as $|\delta F_0|$ is rather small.
Hence, by varying $F_0$ in sufficiently small steps and using a Newton-Raphson
method to adapt the initial condition, we can ``trace'' the periodic orbit as
a function of the field amplitude.

\begin{figure}
\begin{center}
\leavevmode
\epsfxsize8cm
\epsfbox{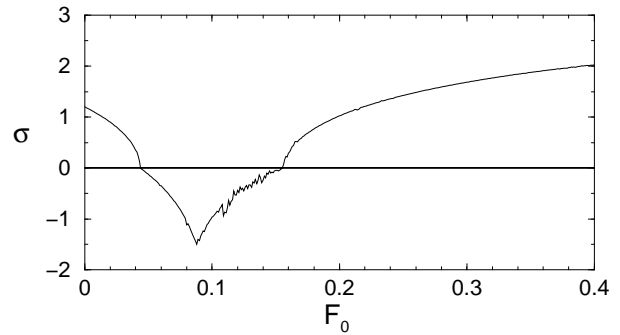}
% page: 600 x 300
\end{center}
\caption{The ``Lyapunov exponent'' $\sigma$ of the asymmetric stretch orbit, i.e.\
  the logarithm of the largest eigenvalue of the stability matrix, is plotted
  as a function of the field amplitude $F_0$. In between $F_0 \simeq 0.044$ and $F_0 \simeq
  0.154$, $\sigma$ is negative, which indicates that the asymmetric stretch orbit
  is stable and constitutes the center of a regular island in the phase space
  of driven helium.
\label{fg:lp}}
\end{figure}

\begin{figure}
\begin{center}
\leavevmode
\epsfxsize8cm
\epsfbox{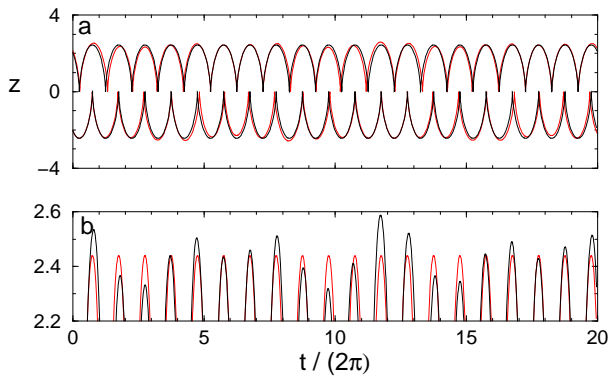}
% page: 600 x 400
\end{center}
\caption{a) The stabilized asymmetric stretch orbit at $F_0 = 0.074$ and $\omega =
  1$, plotted over 20 field cycles.
  Initial condition: $z_1 = - z_2 = 2.0367$, $p_{z1} = p_{z2} = -0.54512$.
  Plotted in grey in the background is a trajectory with a slightly different
  initial condition ($\Delta z_1 = 0.1$).
  The effect of such a deviation from the fundamental periodic orbit is
  magnified in (b) in the vicinity of the upper electron's outer turning
  point (on top of the fundamental orbit which is plotted in grey):
  In contrast to the field-free case (Fig.~\ref{fg:as}), the electrons
  maintain their relative phase and perform stable oscillations around the
  asymmetric stretch orbit.
\label{fg:st}}
\end{figure}

Fig.~\ref{fg:lp} shows the Lyapunov exponent $\sigma = \ln \lambda_{max}$ of the driven
asymmetric stretch orbit, which is calculated from the largest eigenvalue
$\lambda_{max}$ of the stability matrix associated with one oscillation period.
In between $F_0 \simeq 0.044$ and $F_0 \simeq 0.154$, this eigenvalue is smaller than unity,
which formally results in a negative Lyapunov exponent and indicates stable
dynamics in the vicinity of the orbit.
This is indeed confirmed by plotting the corresponding trajectories in
configuration space -- e.g., for $F_0 = 0.074$ as shown in Fig.~\ref{fg:st}:
Instead of increasing exponentially with time, a small deviation from the
periodic orbit leads to stable, quasiperiodic oscillations around the orbit.

A semiclassical estimation of the minimum atomic excitation at which this
regular island supports fully localized quantum states is provided by the
Einstein-Brillouin-Keller (EBK) quantization criterion (see, e.g.,
\cite{Per77ACP,MirKor94JPA}).
The latter states that any quantized torus must fulfill
\begin{equation}
  \oint_{\cal C} {\bf p} d{\bf q} = 2 \pi \hbar \left( n + \frac{\mu}{4} \right) \label{eq:ebk}
\end{equation}
for all closed curves ${\cal C}$ that are contained within the surface of the
torus.
Here, $({\bf p},{\bf q})$ are canonically conjugate phase space variables, 
$n \geq 0$ is an integer, and $\mu$ represents the Maslov index accumulated along
${\cal C}$.
If we choose the curve ${\cal C}$ to encircle the interior of the torus within
the subspace that is spanned by a particular pair $(p_i,q_i)$ of canonically
conjugate variables (which implies $\mu = 2$), we obtain the requirement
that the ``cross section'' area $A_i = \oint p_i d q_i$ enclosed by the torus
within that subspace must equal $2 \pi \hbar ( n + 1/2 )$ for some $n \geq 0$.
Hence, in order to obtain at least one quantized state within the island, we
need to require that its cross section area with respect to any pair
$(p_i,q_i)$ of phase space variables is at least of the order of $\pi \hbar$;
in that case, at least one of the invariant tori would fulfill the EBK
criterion for $n = 0$.

\begin{figure}
\begin{center}
\leavevmode
\epsfxsize8cm
\epsfbox{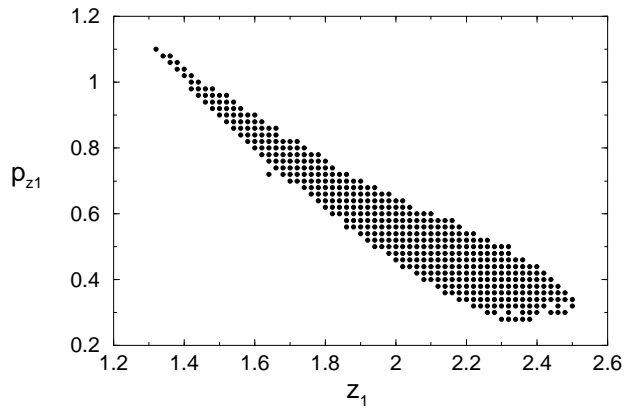}
% page: 600 x 400
\end{center}
\caption{Cross section through the regular island around the stabilized
  asymmetric stretch orbit at $F_0 = 0.074$.
  Plotted are initial conditions in the $z_1$ -- $p_{z1}$ subspace that lead
  to stable quasiperiodic motion after 1000 field cycles, while the initial
  condition of the other electron is fixed to the one of the fundamental
  periodic orbit:
  $z_2 = 2.0367$, $p_{z2} = - 0.54512$.
  $z_1$ and $p_{z1}$ are varied in discrete steps of the size $\Delta z_1 = \Delta
  p_{z1} = 0.02$.
  The cross section area of the island extracted from this figure equals
  $A_0 \simeq 0.179$.
\label{fg:sz}}
\end{figure}

\begin{figure}
\begin{center}
\leavevmode
\epsfxsize8cm
\epsfbox{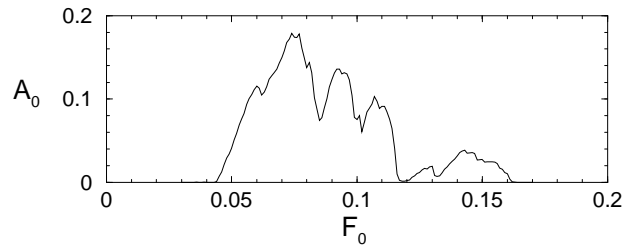}

\end{center}
\caption{Cross section area $A_0$ of the stabilized asymmetric stretch island
  as a function of the field amplitude $F_0$, calculated in the same way as in
  Fig.~\ref{fg:sz}.
  The global maximum is encountered at $F_0 = 0.074$.
\label{fg:sf}}
\end{figure}

Fig.~\ref{fg:sz} visualizes the cross section through the asymmetric stretch
island at $F_0 = 0.074$.
From a discrete lattice of $2867$ equidistant initial conditions within
the range $1.32 \leq z_1 \leq 2.52$ and $0.24 \leq p_{z1} \leq 1.16$ (the initial values of
$z_2$ and $p_{z2}$ are fixed to the ones associated with the fundamental
periodic orbit), only those are plotted that lead to stable quasiperiodic
motion after 1000 field cycles.
From the number $n_s = 447$ of such stable initial points, we obtain the cross
section area of the island within the $z_1$--$p_{z1}$ space (and, by symmetry,
also within the $z_2$--$p_{z2}$ space) according to 
\begin{equation}
A_0 = n_s \Delta z_1 \Delta p_{z1} \simeq 0.179
\end{equation}
with $\Delta z_1 = \Delta p_{z1} = 0.02$ the spacing between adjacent lattice points in
position and momentum, respectively.
This is indeed the largest cross section area of the island that can be
obtained by stabilizing the asymmetric stretch orbit.
Fig.~\ref{fg:sf} shows the cross section area $A_0$, calculated in the above
way, as a function of the field amplitude $F_0$; clearly, the maximum occurs
around $F_0 = 0.074$.

The integrated action along the asymmetric stretch orbit is given by
\begin{equation}
S = \int_0^{2 \pi / \omega} \sum_{i=1,2} p_{zi}(t) \, \frac{d}{dt} z_i(t) \, dt \, . \label{eq:ac}
\end{equation}
From numerical integration, we obtain $S \simeq 19.0 \equiv S_0$ at $\omega = 1$.
If we assume (in analogy to the quantization of Kepler orbits in hydrogen)
that semiclassically $S \simeq 4 \pi N$ for intra-shell asymmetric 
stretch states with principal quantum number $N$ ($\hbar$ is dropped again here
and in the following), and if we take into account that phase space cross
section areas $\oint p dq$ scale in the same way as action integrals when a
scaling transformation according to (\ref {eq:sc}) is performed, we obtain
that a cross section area $A_0$ at the scaling corresponding to $\omega = 1$ is
equivalent to a cross section area $A = 4 \pi N A_0 / S_0$ at the scaling
corresponding to the state with principal quantum number $N$.
Hence, requiring that $A \geq \pi$, we formally obtain the minimum principal
quantum number 
\begin{equation}
N_{min} = S_0 / (4 A_0) \simeq 27 \label{eq:nmin}
\end{equation}
at which fully localized quantum states are to be expected on the stabilized
asymmetric stretch island.

\begin{figure}
\begin{center}
\leavevmode
\epsfxsize8cm
\epsfbox{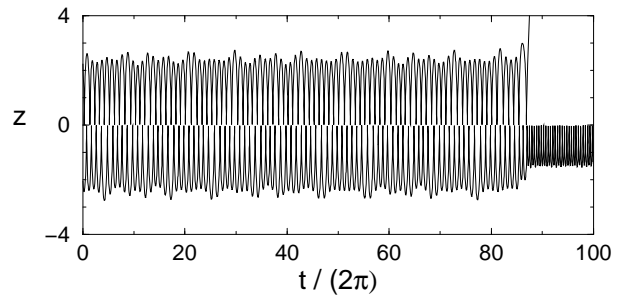}
\end{center}
\caption{Long-time instability of a trajectory starting close to the boundary
  of the regular island at $F_0 = 0.074$.
  The initial conditions are as in Fig.~\ref{fg:st}(a), with the upper
  electron displaced by $\Delta z_1 = 0.2$.
  After a seemingly stable quasi-periodic oscillation over about 85 field
  cycles, the correlation between the electrons ``suddenly'' breaks down.
  A nearby triple collision transfers then a large amount of energy between
  the electrons, which leads to ionization.
\label{fg:tr}}
\end{figure}

It should be noted that the above estimation (\ref{eq:nmin}) is fairly
imprecise and has to be taken as approximate guideline for the order of
magnitude of $N_{min}$, rather than as a precise criterion.
This is, on the one hand, due to the simplifications that are involved in
determining the cross section area of the outermost invariant torus of the
island. 
On the other hand, the notion of an ``outermost'' torus itself is, strictly
speaking, not meaningful in five or more dimensions, where invariant tori
(with codimension larger than unity) do not divide the phase space in
disconnected segments.
As a consequence, chaotic sublayers ``within'' the island are connected with
each other and with the chaotic sea ``outside'' the island, and trajectories
starting on such sublayers can, via Arnold diffusion \cite{LicLie}, leave the
island on finite (though typically rather long) time scales.
This is illustrated in Fig.~\ref{fg:tr}, which shows the time evolution of a
trajectory starting close to the boundary of the island.
After a seemingly stable quasi-periodic oscillation over more than 80
field cycles, the correlation between the electrons breaks down and the atom
ionizes.

At finite values of $\hbar$, moreover, the presence of Cantori
\cite{MacMeiPer84PRL} can inhibit quantum transport also in the immediate
vicinity {\em outside} the regular island (where ``hierarchical states''
are localized \cite{KetO00PRL}). 
The effective size of the localizing region that the quantum system ``sees''
may therefore be considerably larger than the size of the classical island.
Correspondingly, estimations via the EBK criterion are typically rather
conservative and tend to overestimate the actual value of the minimum
scaling needed to obtain a localized state on the island
(an extreme case was reported in \cite{MueBurNoi92PRA,MueBur93PRL} where
quantum states associated with the ``Langmuir'' orbit of helium were
semiclassically predicted for $N > 500$ and quantum mechanically found
at $N = 10$).
Indeed, we shall see in Section \ref{QM} that double excitations with
principal quantum numbers of the order of $N \simeq 10$, which are
experimentally accessible \cite{DomO96PRA,PueO01PRL}, would be sufficient to
obtain nondispersive wave packets anchored on the asymmetric stretch orbit.

\section{Deviations from collinearity}

\label{2D}

In contrast to deviations within the collinear subspace, the {\em transverse}
degrees of freedom do not lead to destabilization of the unperturbed
asymmetric stretch orbit.
If both electron are symmetrically displaced in a direction that is
perpendicular to the $z$ axis -- e.g., by nonvanishing initial components
$y_1(t=0) = y_2(t=0)$) -- the effective repulsion between them is enhanced due
to the reduced screening by the nucleus, which drives the electrons back to
the $z$ axis and leads to stable bending vibrations of the configuration (see
in this context also the ``asynchronous'' orbit discussed in
\cite{GruSim91JPB,Sim97JPBL}).
Marginal stability (with Lyapunov exponent $0$) is encountered for
antisymmetric displacements -- e.g., with $y_1(t=0) = -y_2(t=0)$ -- which is a
consequence of angular symmetry and the resulting conservation of the total
angular momentum.

\begin{figure}
\begin{center}
\leavevmode
\epsfxsize8cm
\epsfbox{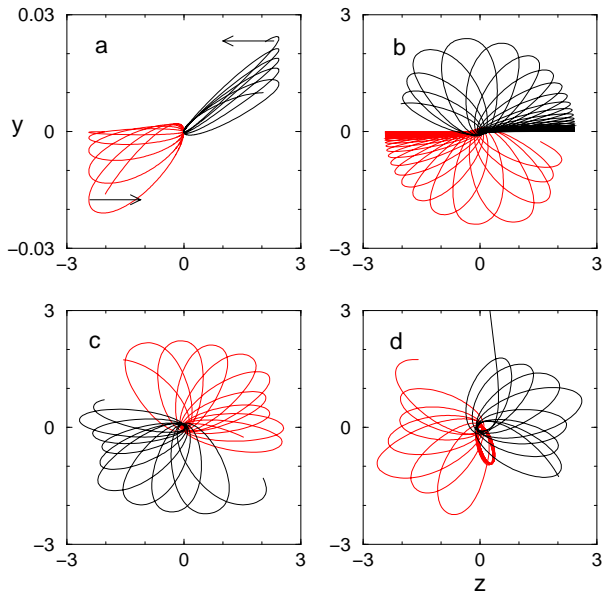}
% page: 600 x 600
\end{center}
\caption{Transverse instability of the driven asymmetric stretch orbit.
  Plotted is, in the $y$--$z$ plane, the time evolution of a trajectory with
  initial condition in the vicinity of the collinear stabilized orbit at 
  $F_0 = 0.074$ (the same initial conditions as in Fig.~\ref{fg:st}(a), with
  the electron on the right-hand side transversally displaced by $\Delta y =
  0.01$).
  The trajectory is shown in the time intervals (a) 0 -- 5 field cycles, (b)
  0 -- 28 field cycles, (c) 28 -- 38 field cycles, and (d) 38 -- 50 field
  cycles.
%  We see that the external field drives the electrons away from the
%  polarization axis (a) (the arrows symbolize the direction of the external
%  force at the outer turning point), which eventually leads to a rotation of
%  the whole configuration around the nucleus (b,c).
%  In the course of this high-dimensional, chaotic motion, the strong
%  correlation between the electrons breaks down (d) and the atom ionizes at
%  about 46 field cycles.
\label{fg:a2}}
\end{figure}

In the presence of an external driving, however, with the phase of the field
chosen such as to stabilize the collinear orbit, the angular symmetry is
broken and the marginally stable degree of freedom is transformed into an
unstable one.
This can be understood from the net torque that the field exerts onto the
atom.
If the configuration is slightly tilted with respect to the $z$ axis (e.g., by
small initial components $y_1(t=0) = -y_2(t=0)$), the external force is no
longer directed parallel to the semimajor axis of the Kepler orbits, but
exhibits a small perpendicular component.
At the outer turning point of the Kepler oscillation, where the contribution
to the net torque is most effective, this perpendicular component of the force
is directed {\em away} from the $z$ axis (see Fig.~\ref{fg:a2}(a)) and
therefore tends to enhance the initially small displacement from collinearity.

This phenomenon of transverse instability arises also in the resonantly driven
hydrogen atom.
In the one-electron case, however, the two-dimensional motion that results
from a small displacement from the field polarization axis is bounded by means
of dynamical barriers in phase space and does not lead to ionization.
Such dynamical barriers do not exist in the two-electron case, due to the high
dimensionality of the phase space.
Fig.~\ref{fg:a2} shows what happens when the configuration starts from an
initial condition that corresponds to the stabilized collinear orbit, with
one of the electrons slightly displaced in $y$ direction.
We clearly see that the external field drives the configuration away from the
$z$ axis (Fig.~\ref{fg:a2}(a)) and rotates it around the nucleus
(Fig.~\ref{fg:a2}(b,c)).
At about 40 field cycles, when one such rotation is completed, the correlation
between the electrons breaks down and the atom eventually undergoes single
ionization.

The scenario encountered here is indeed very similar to the case of the
resonantly driven $Zee$ (frozen planet) configuration
\cite{SchBuc98JPB,SchBuc99PhD,SchBuc03EPD}, which is also unstable against
deviations from collinearity.
For that case as well as for the resonantly driven hydrogen atom
\cite{LeoRic87JPB,SacZakDel98EPD}, it is known that the application of an
additional, static electric field parallel to the polarization of the
driving can enforce transverse stability of the configuration.
For this purpose, the static electric field needs to be directed such that it
prevents the electrons from approaching the $z = 0$ plane -- and thereby
counterbalances the destabilization mechanism induced by the torque of the
time-periodic driving.

\begin{figure}
\begin{center}
\leavevmode
\epsfxsize8cm
\epsfbox{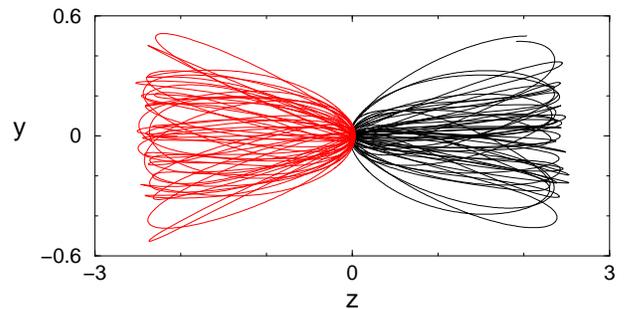}
% page: 600 x 300
\end{center}
\caption{
  Transverse stabilization of the asymmetric stretch orbit at $F_0 = 0.074$ by
  means of a magnetic field $B = 0.5$ polarized along the $z$ axis.
  The initial condition corresponds to the stabilized collinear orbit, with
  the electron on the right-hand side transversally displaced by $\Delta y = 0.5$.
  Plotted are the $y$ and $z$ components of the three-dimensional trajectory
  over 30 field cycles. 
  We see that the transverse confinement provided by the magnetic field leads
  to a stable, quasiperiodic oscillation around the field polarization axis.
\label{fg:b2}}
\end{figure}

In the case of the driven $eZe$ configuration, a homogeneous static electric
field cannot be used to prevent transverse destabilizaton, since it cannot be
oriented in such a way that it simultaneously keeps both electrons from
approaching the $z=0$ plane.
An effective alternative, however, is provided by the possibiliy of adding a
strong static {\em magnetic} field, which is oriented along the $z$ axis,
i.e., parallel to the polarization of the driving.
Due to the tight diamagnetic confinement created by such a magnetic field,
small deviations from collinearity are not enhanced, but lead to stable
(three-dimensional) rotations of the electrons around the $z$ axis.
This is illustrated in Fig.~\ref{fg:b2} which shows the two-dimensional
projection of a trajectory that starts in the vicinity of the collinear
stabilized asymmetric stretch orbit at $F_0 = 0.074$, in the presence of a
magnetic field with strength $B_0 = 0.5$ oriented along the $z$ axis.
The initial displacement of the electron on the right-hand side is $y_1 =
0.5$, which indicates a rather large phase space volume of the associated
regular island.

For the asymmetric stretch orbit at $F_0 = 0.074$, we find that magnetic field
strengths in the range $B_0 \simeq 0.2 \ldots 2$ can be used to provide transverse
stability of the configuration \cite{largeB}.
Since a homogeneous magnetic field ${\bf B}$ is incorporated into the
Hamiltonian via the substitution
\begin{equation}
{\bf p}_i \longmapsto {\bf p}_i + \frac{1}{2} ( {\bf B} \times {\bf r} )
\end{equation}
($i = 1,2$), it is scaled according to ${\bf B} \longmapsto \nu^{-3} \, {\bf B}$
when a scaling transformation of the type (\ref{eq:sc}) is applied.
Hence, at double excitations corresponding to the principal quantum number 
$N = 25 \ldots 30$, where the EBK criterion predicts fully localized quantum states
on the asymmetric stretch island, a ``scaled'' magnetic field strength 
$B_0 = 0.2$ (at $\omega = 1$) would roughly correspond to 
$B = ( 4 \pi N / S_0 )^{-3} B_0 \sim 10$~Tesla (with $S_0 \simeq 19$, see Section \ref{CL}).
Similarly, the resonant driving frequency would be given by 
$\omega = ( 4 \pi N / S_0 )^{-3} \sim 2 \pi \times 10^{12}$ Hz.
To realize the field amplitude $F = ( 4 \pi N / S_0 )^{-4} F_0$ with $F_0 \simeq 0.07$, 
radiation intensities of the order of $I \sim 10^5$ W$/$cm$^2$ would be required.
This points towards quantum cascade lasers as a possible source for the
electromagnetic radiation \cite{Koe02N}.

\section{Quantum signatures of the stabilized asymmetric stretch orbit}

\label{QM}

After having shown the possibility to stabilize the asymmetric stretch orbit
in the classical $eZe$ configuration of helium, we discuss in this
section to which extent this classical stabilization manifests itself in the
corresponding quantum system.
In principle, a full-blown three-dimensional treatment of doubly excited
helium in the electromagnetic field would be desirable in this context.
While such {\it ab initio} calculations can indeed be performed in the
unperturbed two-electron atom (up to principal quantum numbers of the order of
$N \simeq 15 \ldots 20$ \cite{GreDel97EPL} -- which, as we shall see later on, should be
sufficient for our purpose), they are still impractical in the presence of a
nonperturbative time-dependent electric field where the conservation of the
total angular momentum is broken and the Floquet formalism is required to
obtain wave packet states associated with the resonantly driven collinear
orbit.
Essential properties of these wave packets, however, can be reproduced from a
one-dimensional model of the two-electron atom which properly takes into
account the dynamics along the field polarization axis.
This is particularly the case if a magnetic field is applied in order to
provide stability in the transverse degrees of freedom;
the electrons are then strongly confined in the $x$--$y$ plane and evolve
according to an effective one-dimensional dynamics along the $z$ axis.

To correctly describe this dynamics, we demand that the one-dimensional model
represents the exact quantum analog of the classical $eZe$ configuration.
This in particular requires to take into account the full Coulomb interaction
between the charged particles.
A smoothening of the Coulomb singularity, which is frequently employed in
one-dimensional models of driven atoms (e.g., \cite{GroEbe92PRL}), may not be
permitted here, since it would lead to a considerably different behaviour in
the corresponding classical system.

Consequently, we write the Hamiltonian that generates the quantum dynamics of
the driven collinear configuration as
\begin{eqnarray}
H & = & - \frac{1}{2} \frac{\partial^2}{\partial \zeta_1^2} 
  \; - \; \frac{1}{2} \frac{\partial^2}{\partial \zeta_2^2} 
  \; - \; \frac{Z}{\zeta_1} \; - \; \frac{Z}{\zeta_2} \; + \; \frac{1}{\zeta_1 + \zeta_2} 
  \nonumber \\
&& - \frac{F}{{\rm i} \, \omega} \, \sin \omega t \left( \frac{\partial}{\partial \zeta_1}  -  
  \frac{\partial}{\partial \zeta_2} \right) \label{eq:hqm}
\end{eqnarray}
with $\zeta_1 = z_1$ and $\zeta_2 = - z_2$ the absolute values of the (Cartesian)
coordinates of the electrons along the field polarization axis.
The external field, parametrized by the amplitude $F$ and the frequency $\omega$,
is incorporated in the velocity gauge in order to ensure good convergence of
the numerical calculation \cite{Sha88ZPD}.
Effectively, the electrons appear here as {\em distinguishable} particles
(with $\zeta_1,\zeta_2 > 0$), which is consistent with the classical impenetrability of
the Coulomb singularity at the origin.

Due to the temporal periodicity of the Hamiltonian, the Schr{\"o}dinger
problem represented by (\ref{eq:hqm}) is conveniently treated in the framework
of Floquet theory.
The latter states that any solution $\psi_t(\zeta_1,\zeta_2)$ of the Schr{\"o}dinger equation
can be written as
\begin{equation}
  \psi_t(\zeta_1,\zeta_2) = \int\hspace*{-0.5cm}\sum d {\mathcal E} \, C_{\mathcal E} \, 
  \psi_t^{({\mathcal E})}(\zeta_1,\zeta_2) \, e^{- {\rm i} {\mathcal E} \, t }
\end{equation}
with time-independent complex expansion coefficients $C_{\mathcal E}$,
where ${\mathcal E}$ are the quasienergies and $\psi_t^{({\mathcal E})}(\zeta_1,\zeta_2) = 
\psi_{t+ 2 \pi / \omega}^{({\mathcal E})}(\zeta_1,\zeta_2)$ the associated time-periodic
quasienergy-eigenfunctions.
The latter are determined by the time-independent eigenvalue equations
\begin{eqnarray}
  (H_0 \; + \; k \omega \; - \; {\mathcal E}) \, \hat{\psi}_k^{(\mathcal E)}(\zeta_1,\zeta_2) &&
  \nonumber \\
  + \; V \, ( \hat{\psi}_{k+1}^{(\mathcal E)}(\zeta_1,\zeta_2) \; - \; \hat{\psi}_{k-1}^{(\mathcal
    E)}(\zeta_1,\zeta_2) ) & = & 0
\end{eqnarray}
with
\begin{eqnarray}
H_0 & = & - \frac{1}{2} \frac{\partial^2}{\partial \zeta_1^2} 
  \: - \: \frac{1}{2} \frac{\partial^2}{\partial \zeta_2^2} 
  \: - \: \frac{Z}{\zeta_1} \: - \: \frac{Z}{\zeta_2} \: + \: \frac{1}{\zeta_1 + \zeta_2} \, ,
  \nonumber \\
V & = & \frac{F}{2 \omega} \left( \frac{\partial}{\partial \zeta_1} - \frac{\partial}{\partial \zeta_2} \right) \, ,
\end{eqnarray}
which result from the Fourier series expansion 
\begin{equation}
  \psi_t^{({\mathcal E})}(\zeta_1,\zeta_2) = \sum_{k=-\infty}^{\infty} \hat{\psi}_k^{({\mathcal
      E})}(\zeta_1,\zeta_2) \,
  e^{{\rm i} \, k \omega t} \, .
\end{equation}

For atomic systems, the $\omega$-periodic Floquet spectrum of quasienergies
${\mathcal E}$ is absolutely continuous:
each bound state of the unperturbed atom is coupled to the atomic 
continuum via multiphoton transitions and therefore appears as a
resonance structure in the spectrum.
In order to separate these spectral resonances from the flat background,
we employ the method of complex scaling (see \cite{Moi98PR} for a recent
review), which has proven its usefulness in a variety of numerical 
studies in atomic and molecular physics.
Essentially, the electronic coordinates are complexified according to 
$\zeta_j \mapsto \zeta_j e^{{\rm i} \theta}$, and the momenta according to $\partial / \partial \zeta_j \mapsto 
e^{-{\rm i} \theta} \partial / \partial \zeta_j$  ($j = 1, 2$).
Formally, this scaling can be expressed by the (non-unitary) transformation
\begin{equation}
  \psi_t^{({\mathcal E})}(\zeta_1,\zeta_2) \; \longmapsto \; e^{{\rm i} \theta} \, \psi_t^{({\mathcal E})}(\zeta_1
  e^{{\rm i} \theta}, \zeta_2 e^{{\rm i} \theta})
\end{equation}
of the quasienergy-eigenfunctions, as well as by the corresponding
transformation
\begin{eqnarray}
  && H \left( \zeta_1, \zeta_2, \frac{\partial}{\partial \zeta_1},\frac{\partial}{\partial \zeta_2}, t \right) \; \longmapsto 
  \nonumber \\
  && H \left( \zeta_1 e^{{\rm i} \theta}, \zeta_2 e^{{\rm i} \theta}, 
     e^{-{\rm i} \theta} \frac{\partial}{\partial \zeta_1}, e^{-{\rm i} \theta} \frac{\partial}{\partial \zeta_2}, t \right)
\end{eqnarray}
of the Hamiltonian.
As a result, we obtain a complex symmetric rather than Hermitian eigenvalue
problem.
Resonance structures in the continuous spectrum of the ``real'', unrotated
eigenvalue problem ($\theta = 0$) appear now as discrete complex eigenvalues
${\mathcal E} = E - {\rm i} \Gamma/2$, with the real and imaginary parts
corresponding to the energies $E$ and spectral widths $\Gamma/2$ (HWHM -- the half
width at half maximum) of the resonances, respectively.

In perfect analogy to the quantum description of the driven $Zee$
configuration of helium (see \cite{SchBuc03EPD} for more details), the complex
scaled Floquet Hamiltonian is now expanded in the product basis
\begin{equation}
\{ S_n^{(\alpha)}(\zeta_1) S_m^{(\alpha)}(\zeta_2)\, : n,m \geq 1 \} \label{eq:sturm2}
\end{equation}
composed of the real-valued Sturmian functions
\begin{equation}
  S_n^{(\alpha)} ( \zeta ) \; = \; \frac{(-1)^n}{\sqrt{n}} \, \frac{2 \zeta}{\alpha} \,  
  \exp \left( - \frac{\zeta}{\alpha} \right) L_{n - 1}^{(1)} \left( \frac{2 \zeta}{\alpha} \right),
  \label{eq:sturm}
\end{equation}
where the $L_n^{(l)}$ denote the associated Laguerre polynomials.
By the choice of this basis, the Hilbert space is effectively restricted to
functions that scale at least linearly with $\zeta_j$ for small $\zeta_j$ ($j=1,2$)
and therefore do not exhibit divergent potential matrix elements.
The scaling parameter $\alpha$, which can be freely chosen in principle, determines
the maximum spatial extension (as well as the minimum coarse graining) that is
represented by a truncated basis set  $\{S_1^{(\alpha)}, \ldots, S_N^{(\alpha)} \}$.
Variations of $\alpha$ (as well as of the complex scaling angle $\theta$) provide
an efficient means to verify the numerical convergence of the calculation.

In the unperturbed atomic system ($F = 0$), the eigenstates of $H$ are
characterized by a well-defined ``parity'': they are either ``even'' or
``odd'' with respect to the exchange of $\zeta_1$ and $\zeta_2$:
\begin{equation}
\psi(\zeta_1,\zeta_2) = \pm \psi(\zeta_2,\zeta_1) \, .
\end{equation}
Within each of these two symmetry classes, the spectrum can be grouped into
series that converge towards single ionization thresholds, which are labelled
by the principal quantum number $N$ of the inner electron.
Naturally, this classification cannot be carried out in a perfectly rigorous
way, since $N$ is not a good quantum number.
In practice, ambiguities arise as soon as states belonging to different series
start to resonantly interact with each other, which is the case for $N \geq 5$
(see also \cite{BluRei91}).

\begin{figure}
\begin{center}
\leavevmode
\epsfxsize8cm
\epsfbox{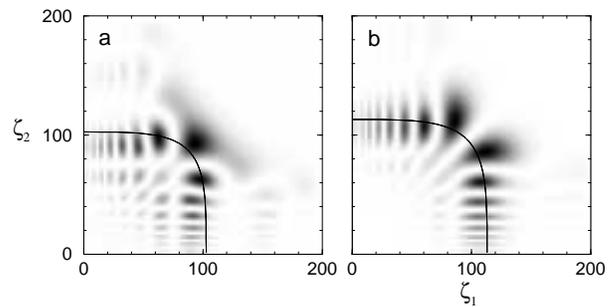}
% page: 600 x 300
\end{center}
\caption{Probability densities of asymmetric stretch states in configuration
  space.
  Plotted are in (a) the energetically lowest ``even'' and in (b) the
  energetically lowest ``odd'' state at principal quantum number $N = 10$.
  We clearly see a localization along the classical asymmetric stretch orbit,
  which is shown by the underlying curve.
  With respect to Fig.~\ref{fg:as}(a), the orbit was scaled according to
  $\zeta_j \mapsto (2 \pi n_{\rm node} / S_0)^2 \zeta_j$ ($j,2$), with the number of nodes along
  the orbit given by $n_{\rm node} = 20$ for the even and $n_{\rm node} = 21$
  for the odd state. 
  \label{fg:aq}}
\end{figure}

For each parity, the energetically lowest member of each series exhibits a
pronounced localization in the vicinity of the classical asymmetric stretch
orbit.
This is illustrated in Fig.~\ref{fg:aq}, which shows the densities
of the lowest even and odd states at principal quantum number $N = 10$.
The probability distributions of these ``asymmetric stretch states'' are in
fact rather similar to the projections of three-dimensional intra-shell
eigenfunctions (with maximum interelectronic angle) onto the $\zeta_1$ -- $\zeta_2$
subspace \cite{RicWin93JPB}, which are also localized along the collinear 
asymmetric stretch orbit.

\begin{figure}
\begin{center}
\leavevmode
\epsfxsize8cm
\epsfbox{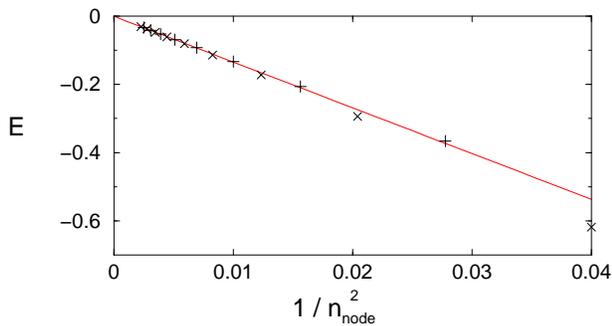}
% page: 600 x 300
\end{center}
\caption{Energies of the asymmetric stretch states, as a function of $1 /
  n_{node}^2$ with $n_{node}$ the number of nodes along the asymmetric stretch
  orbit. 
  The signs $+$ and $\times$ denote ``even'' and ``odd'' states, respectively.
  The grey line shows the scaling $E = - 13.4 / n_{\rm node}^2$, which is
  expected in the semiclassical limit.
\label{fg:e0}}
\end{figure}

Due to the enforced node at $\zeta_1 = \zeta_2$, the energy $E_{N-}$ of the odd asymmetric
stretch state is, at given $N$, considerably enhanced with respect to the
energy $E_{N+}$ of the even state, and lies approximately in the middle
between $E_{N+}$ and $E_{(N+1)+}$.
More precisely, the scaling laws (\ref{eq:sc}) predict the relation 
$E \simeq E_0 ( S / S_0 )^{-2}$ between the energy $E$ and the action $S$
(as defined by (\ref{eq:ac})) in the semiclassical limit, where $E_0 \simeq - 1.50$
and $S_0 \simeq 18.8$ are the respective values of the energy and action at
frequency $\omega = 1$ (for the unperturbed asymmetric stretch orbit).
Semiclassical quantization of the action along the orbit yields 
$S \simeq 2 \pi n_{\rm node}$ for the asymmetric stretch states. 
Here, $n_{\rm node}$ denotes the number of nodes of the wavefunction along the
orbit (including the nodes at $\zeta_1=0$ and $\zeta_2=0$), which is related to the 
principal quantum number $N$ according to
\begin{equation}
n_{\rm node} = \left\{ \begin{array}{r@{\quad}l} 2 N & \mbox{for even states}
    \\ 2 N + 1 & \mbox{for odd states} \end{array} \right. \, .
\end{equation}
The resulting scaling 
\begin{equation}
E_{N\pm}[{\rm a.u.}] \simeq E_0 ( 2 \pi n_{\rm node} / S_0 )^{-2} 
\simeq - 13.4 / n_{\rm node}^2 \label{eq:esc}
\end{equation}
is, as shown in Fig.~\ref{fg:e0}, indeed encountered in the numerically
calculated spectrum.

%\begin{figure}
%\begin{center}
%\leavevmode
%\epsfxsize8cm
%\epsfbox{enfl.eps}
%% page: 600 x 200
%\end{center}
%\caption{Energies of the unperturbed asymmetric stretch states, shifted by
%  integer multiples of $\omega = (16 \pi / S_0)^{-3} \simeq 0.00653$ a.u.\ --- i.e.,
%  $E_{N\pm} \mapsto E_{N\pm} - k \omega$ --- into the Floquet zone $-0.05653$ a.u. $\leq E \leq
%  -0.056$ a.u.
%  Within $14 \leq n_{\rm node} \leq 18$, the shift integer $k$ differs by one
%  between adjacent states (more precisely, $k = n_{\rm node} - 16$ in this
%  range), which means that these states are strongly coupled to each other via
%  ``single photon'' transitions.
%  As soon as the perturbation is switched on, the even $N = 8$ state (with
%  $n_{\rm node} = 16$) is shifted towards higher energies due to the repulsion
%  from the odd states at $N = 8$ and $N = 9$ (with $n_{\rm node} = 15$ and
%  $17$, respectively), and eventually undergoes a diabatic transition into a
%  nondispersive wave packet localized on the stabilized asymmetric stretch
%  orbit.
%\label{fg:ef}}
%\end{figure}

For a resonant stabilization of the asymmetric stretch orbit associated with
the state with node number $n_{\rm node}$, the driving frequency needs to be
set to $\omega = \nu^{-3}_{n_{\rm node}}$ where $\nu_{n_{\rm node}}$ denotes the
scaling parameter in (\ref{eq:sc}) that generates the transformation
(\ref{eq:sc}) from the ``scaled'' orbit (at $\omega = 1$) to the actual orbit
underlying this state.
Semiclassically, the action of the lowest quantized state within the
stabilized asymmetric stretch island is given by
\begin{equation}
S \simeq 2 \pi ( n_{\rm node} + (\gamma_1 + \gamma_2) / 2 )
\end{equation}
where $\gamma_1$ and $\gamma_2$ denote the winding numbers along the stabilized orbit
(compare, e.g., with \cite{WinRicTan92CHAOS}).
By virtue of $S = S_0 \nu_{n_{\rm node}}$, therefore, we infer
\begin{eqnarray}
  \nu_{n_{\rm node}} & = & 2 \pi ( n_{\rm node} + (\gamma_1 + \gamma_2) / 2 ) / S_0 \nonumber \\
  & \simeq & 0.33 ( n_{\rm node} + 0.215 ) \, ,
\label{eq:scq}
\end{eqnarray}
as evaluated for the numerically calculated values $S_0 \simeq 19.0$, $\gamma_1 \simeq
0.174$, and $\gamma_2 \simeq 0.256$ (as obtained from the diagonalization of the
monodromy matrix) at $F_0 = 0.074$.
Comparison of the resulting approximate scaling $\omega \sim 28 / n_{\rm node}^3$ with
(\ref{eq:esc}) shows that $\omega$ roughly corresponds to the level spacing between
two consecutive asymmetric stretch states.
This implies that these states are nearly degenerate in the resulting Floquet
spectrum, and become strongly coupled to each other as soon as the
time-periodic perturbation is switched on.
Provided the size of the classical island is large enough, this coupling gives
rise to a Floquet state that corresponds to a nondispersive two-electron wave
packet anchored on the stabilized asymmetric stretch orbit.

\begin{figure}
\begin{center}
\leavevmode
\epsfxsize8cm
\epsfbox{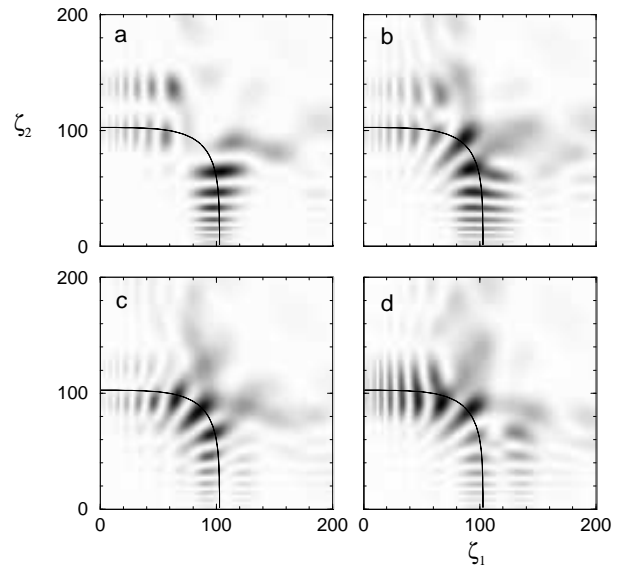}
% page: 600 x 600
\end{center}
\caption{Nondispersive two-electron wave packet anchored on the stabilized
  asymmetric stretch orbit.
  Plotted are the probability densities $|\psi_t(\zeta_1,\zeta_2)|^2$ of the Floquet
  state $\psi_t$ at (a) $\omega t = 0$, (b) $\omega t = \pi / 4$, (c) $\omega t = \pi / 2$, 
  (d) $\omega t = 3 \pi / 4$.
  The driving frequency is set to $\omega = 0.0034$ a.u.\ $\simeq \nu_{20}^{-3}$
  (see Eq.~(\ref{eq:scq})), i.e., the wave packet is centered around
  the even asymmetric stretch state at $N = 10$.
  Accordingly, the field amplitude was chosen as $F = 3.7 \times 10^{-5}$ a.u.\ 
  $\simeq \nu_{20}^{-4} F_0$ with $F_0 = 0.074$
  (the underlying orbit was accordingly scaled, see Fig.~\ref{fg:aq}(a)).
  Significant contributions to the probability density far from the orbit
  reveal that the classical resonance island is, at this value of $N$, not
  yet big enough to support a fully localized quantum state.
  \label{fg:wp}}
\end{figure}

Contrary to the estimation based on the EBK criterion (see Section \ref{CL}),
such nondispersive wave packets are already found at principal quantum numbers
$N \geq 8$.
An example of such a wave packet, centered around the even asymmetric stretch
state at $N = 10$, is shown in Fig.~\ref{fg:wp}, where the probability density
of the corresponding Floquet state is plotted for the driving phases $\omega t =
0$, $\pi/4$, $\pi/2$, and $3\pi / 4$.
The driving frequency was set to $\omega = 0.0034$ a.u.\ $\simeq \nu_{20}^{-3}$,
and the field amplitude was accordingly chosen as $F = 3.7 \times 10^{-5}$ a.u.\ 
$\simeq \nu_{20}^{-4} F_0$, with $F_0 =0.074$ the scaled amplitude at which the size of
the classical island is maximized.
We clearly see that the wave packet faithfully traces the motion of the
classical trajectory along the asymmetric stretch orbit.
Significant contributions to the probability density far from the orbit
indicate, however, that the Floquet state is not yet perfectly localized on
the island at this particular value of $N$.

\begin{figure}
\begin{center}
\leavevmode
\epsfxsize8cm
\epsfbox{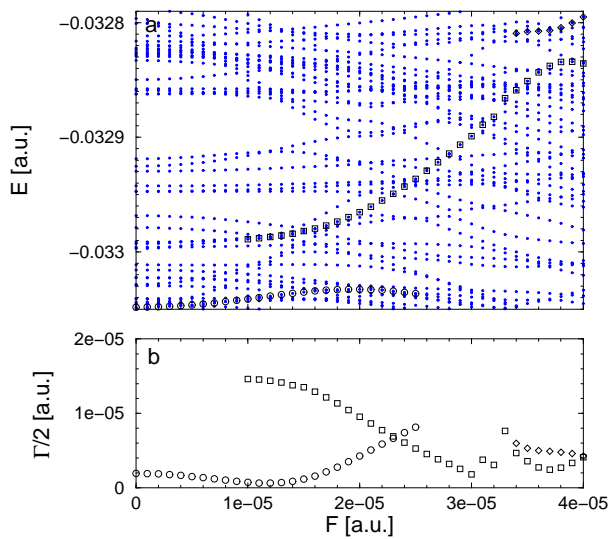}
% page: 600 x 600
\end{center}
\caption{Evolution of the nondispersive wave packet state in the Floquet
  spectrum.
  The upper graph shows the quasienergies of the driven collinear atom at 
  $\omega = 0.0034$ a.u.\ as a function of the field amplitude $F$.
  Plotted are the levels of resonances with ionization rate $\Gamma/2 < 10^{-4}$
  that are coupled to the asymmetric stretch state by at most 10 photons.
  The marks $\circ$, $\scriptstyle \Box$ and $\diamond$ indicate the Floquet states that
  share the contributions of the wave packet state associated with the
  stabilized asymmetric stretch orbit.
  The corresponding ionization rates (HWHM) of these states are shown in (b).
  Indeed, a local minimum of $\Gamma/2$, ``spoiled'' by near-resonant admixtures of
  other, more unstable components, is encountered in the range $F \simeq 3 \times
  10^{-5} \ldots 4 \times 10^{-5}$ a.u. where the size of the classical resonance island
  becomes maximal. 
\label{fg:ew}}
\end{figure}

The spectral evolution from the unperturbed asymmetric stretch state to the
nondispersive wave packet on the stabilized orbit is illustrated in
Fig.~\ref{fg:ew}(a) which shows the Floquet spectrum as a function of the
field amplitude $F$.
Clearly, the Floquet level associated with the stabilized asymmetric stretch
orbit is shifted towards higher quasienergies with increasing $F$ (see
\cite{SchBuc03EPD} for the analogous case in $Zee$ helium), which is in
accordance with the fact that the energy of the classical orbit becomes
slightly enhanced as well.
Note the occurence of huge avoided crossings (e.g., between the $\circ$ and the
$\scriptstyle \Box$ states), which make it difficult to determine the states that
contain significant contributions to the driven asymmetric stretch state
(in practice, the states were identified by a visual inspection of the
probability density in configuration space).

The corresponding ionization rates $\Gamma/2$ (HWHM) of these Floquet states are
shown in Fig.~\ref{fg:ew}(b).
Despite appreciable admixtures of other, less stable components around $F \simeq
3.3 \times 10^{-5}$ a.u.\ which ``spoil'' the ionization rate, it becomes apparent
that a local minimum of $\Gamma/2$ occurs in the range $3 \times 10^{-5}$ a.u.\ $< F < 4
\times 10^{-5}$ a.u.\ where the cross section area of the classical island is
maximized.
At $F = 3.7 \times 10^{-5}$ a.u.\ in particular, the ionization rate $\Gamma/2 \simeq 2.5 \times
10^{-6}$ a.u.\ is obtained for the stabilized asymmetric stretch state.
This implies that a wave packet that is initially prepared on the Floquet
state shown in Fig.~\ref{fg:wp} propagates, without spreading, along the
asymmetric stretch orbit for about 5000 field cycles, until the population is
completely distributed in the ionization continuum.

As is clearly shown in Fig.~\ref{fg:ew}(b), the minimal ionization rate 
in the range $3 \times 10^{-5}$ a.u.\ $< F < 4 \times 10^{-5}$ a.u.\ is of the same
order as the width of the unperturbed asymmetric stretch state.
Globally, therefore, the external driving does not lead to a significant
stabilization at $N = 10$, which can be traced back to the fact that the wave
packet cannot be fully contained within the classical resonance island (see
Fig.~\ref{fg:wp}).
A more pronounced stabilization effect is expected in the regime $N \simeq 20 \ldots 30$
for which the EBK quantization criterion predicts the existence of fully
localized quantum states on the resonance island.
Such quasi-bound states decay only via tunneling-induced couplings to the
surrounding chaotic sea and to the ionization continuum, and should therefore
exhibit significantly lower ionization rates than the unperturbed asymmetric
stretch states.
Calculations in this regime, which require a substantially higher numerical
effort than for $N = 10$, are presently under way.

\section{Conclusion}

In summary, we have shown that the collinear asymmetric stretch orbit of the
classical helium atom can be stabilized by means of a resonant electromagnetic
perturbation.
Within a reasonable range of field amplitudes ($0.04 
{\scriptstyle {< \atop ^\sim}} F_0 {\scriptstyle {< \atop ^\sim}} 0.15$
at $\omega = 1$), the external driving induces a local regular island in the phase
space of collinear motion, in which the electrons perform stable quasiperiodic
oscillations around the stabilized asymmetric stretch orbit.
The driven collinear atom is unstable with respect to deviations from
collinearity.
However, a static magnetic field, oriented parallel to the driving field
polarization, can be used to stabilize the configuration by transversally
confining the electrons to the vicinity of the field polarization axis.

The quantum signatures of this classical stabilization phenomenon were
elucidated by means of a one-dimensional model of the two-electron atom which
captures the essential properties of the dynamics along the field polarization
axis.
Within the Floquet spectrum of the driven atom, wave packet states were
identified which follow, without spreading, the time-periodic Kepler-like
motion along the stabilized asymmetric stretch orbit.
Such nondispersive wave packets, which constitute the quantum analog of the
classical nonlinear resonance, arise for symmetric double excitations with
principal quantum number $N \geq 8$.
At $N = 10$, a local maximum of the life time of the order of 5000 field
cycles was found near the scaled field amplitude $F_0 \simeq 0.07$ where the cross
section area of the classical resonance island is maximized.

The experimental realization of such nondispersive wave packets is certainly
difficult, but does not seem impossible.
Intrashell ``asymmetric stretch''-like states with maximum interelectronic
angle were already populated in high-resolution photoionization experiments
up to the principal quantum number $N = 10$ \cite{DomO96PRA,PueO01PRL}.
At that double excitation, the resonant driving of the asymmetric stretch
orbit would correspond to a laser frequency of the order of $\omega / (2 \pi) \sim
10^{13}$ Hz, while a laser intensity of the order of $I \sim 10^8$ W$/$cm$^2$
would be required for stabilization.
On the other hand, the magnetic field that is necessary to enforce the
transverse confinement along the field polarization axis would exceed $100$
Tesla at $N = 10$, which is too large to be realized in table-top like
experimental setups.
It is not excluded, however, that wave packet states can be localized on the
driven collinear orbit even {\em without} the presence of an additional
perturbation that stabilizes the dynamics in the transverse degrees of freedom
(in the same way as ``scars'', which are anchored on unstable periodic orbits
in chaotic systems \cite{Hel84PRL}).
To examine to which extent such ``scarred'' wave packet states exist in the
absence of a stabilizing magnetic field (and to which extent the classically
allowed decay via the transverse degrees of freedom would influence their life
time), a more quantitative study, taking into account the dynamics within the
two-dimensional configuration space, would be required.

\begin{acknowledgments}
We thank A.~Buchleitner, L.~B.~Madsen, B.~P{\"u}ttner, K.~F.~Renk, and K.~Richter
for useful and inspiring discussions. D.P.\ and P.S.\ gratefully acknowledge
financial support by the Deutsche Forschungsgemeinschaft.
\end{acknowledgments}

\appendix

\section{Numerical integration of the classical equations of motion}

\label{ap:ks}

In this Appendix, we describe the canonical ``Kustaanheimo-Stiefel''
transformation \cite{KusSti65JRAM} which is employed in order to ensure stable
numerical integration of Hamilton's equations of motion including
electron-nucleus collisions.
We restrict the description to the special case of collinear motion of the
electrons in the $eZe$ arrangement.
Generalizations to two- or three-dimensional dynamics can be found in Refs.
\cite{RicTanWin93PRA} and \cite{SchBuc99PhD}, respectively.

In the subspace of collinear $eZe$ motion, the Hamiltonian of driven helium is
given by
\begin{equation}
H = \frac{p_{z1}^2}{2} + \frac{p_{z2}^2}{2} - \frac{Z}{z_1} + \frac{Z}{z_2} +
\frac{1}{z_1 - z_2} + (z_1 + z_2) F \cos \omega t \label{eq:h1d}
\end{equation}
with $z_1 > 0$ and $z_2 < 0$.
Obviously, a direct numerical integration of the equations of motion resulting
from (\ref{eq:h1d}) leads to a diverging momentum $p_{zi} \sim z_i^{-1/2}$ ($i = 1,2$)
whenever one of the electrons hits the nucleus.
We therefore perform the canonical transformation
\begin{equation}
\begin{array}{c@{\;\mapsto\;}c@{\;\equiv\;}c@{\qquad}c@{\;\mapsto\;}c@{\;\equiv\;}c}
\displaystyle
z_1 & Q_1 & \sqrt{z_1}, & p_{z1} & P_1 & 2 \sqrt{z_1} \, p_{z1}, \\
z_2 & Q_2 & \sqrt{-z_2}, & p_{z2} & P_2 & 2 \sqrt{-z_2} \, p_{z2},
\end{array}
\end{equation}
which is associated with the generating function
\begin{equation}
F(p_{z1},p_{z2},Q_1,Q_2) = p_{z1} Q_1^2 + p_{z2} Q_2^2.
\end{equation}
By construction, these new momentum variables $P_1$, $P_2$ remain finite at
electron-nucleus collisions.

In addition to this canonical transformation, we introduce the new effective
Hamiltonian
\begin{eqnarray}
{\mathcal H} & = & {\mathcal H}(P_1,P_2,Q_1,Q_2,E,t) \nonumber \\
& \equiv & Q_1^2 \, Q_2^2 \, (H(p_{z1},p_{z2},z_1,z_2,t) - E) \nonumber \\
& = & \frac{1}{8}( Q_2^2 P_1^2 + Q_1^2 P_2^2 ) - Z ( Q_1^2 + Q_2^2 )  - E
Q_1^2 Q_2^2 \nonumber \\
& & + \frac{Q_1^2 Q_2^2}{Q_1^2 + Q_2^2} + Q_1^2 Q_2^2 ( Q_1^2 - Q_2^2 ) F \cos \omega t
\end{eqnarray}
which defines, via the Hamiltonian equations of motion,
\begin{eqnarray}
\frac{d P_i}{d \tau} & = & - \frac{\partial {\mathcal H}}{\partial Q_i} \label{eq:Ptau} \\
\frac{d Q_i}{d \tau} & = & \frac{\partial {\mathcal H}}{\partial P_i} 
\end{eqnarray}
the time evolution of $P_1$, $P_2$, $Q_1$, $Q_2$ with respect to the pseudo
time $\tau$.
Note that $t$ and $E$ appear as additional canonically conjugate variables,
satisfying
\begin{eqnarray}
\frac{d E}{d \tau} & = & \frac{\partial {\mathcal H}}{\partial t} 
\; = \; Q_1^2 \, Q_2^2 \, \frac{\partial H}{\partial t} \\
\frac{d t}{d \tau} & = & - \frac{\partial {\mathcal H}}{\partial E} \; = \; Q_1^2 \, Q_2^2 
\label{eq:ttau}
\end{eqnarray}
Indeed, one can straightforwardly show that the combined set of differential
equations (\ref{eq:Ptau}--\ref{eq:ttau}) leads to the same time evolution of
$z_i$ and $p_i$ as the original equations of motion derived from (\ref{eq:h1d}),
provided the initial value of the energy variable is set to $E = H|_{t=0}$.
Eqs.~(\ref{eq:Ptau}--\ref{eq:ttau}) permit stable numerical integration over
simple electron-nucleus collisions, and become singular only in the
exceptional event of a triple collision.

\section{Identification of the periodically driven asymmetric-stretch orbit}

\label{ap:po}

The periodic orbits of the driven $eZe$ configuration of helium are identified
using the so-called stability transformation (ST) method 
\cite{SchDia97PRL,DiaSchBih98PRL,SchDia98PRE,PinO00PRE}.
The latter allows to detect unstable periodic orbits of length $p$ in
the $n$-dimensional chaotic time-discrete systems 
\begin{equation}
\mathbf{f} : \mathbf{X} \longmapsto \mathbf{X}' = \mathbf{f}(\mathbf{X}) \, ,
\end{equation}
i.e. the fixed points of the $p$ times iterated map
$\mathbf{X} \longmapsto \mathbf{f}^{(p)}(\mathbf{X})$. 

To this end, a stability transformed system $\mathbf{s}$ is introduced
according to 
\begin{equation}
%\mathbf{s}:~~~\mathbf{X}_{j+1}=\mathbf{s}(\mathbf{X}_j)=\mathbf{X}_j+\lambda\mathbf{C}
\mathbf{s}:\mathbf{X} \longmapsto \mathbf{s}(\mathbf{X}) = \mathbf{X}+\lambda\mathbf{C}
[\mathbf{f}^{(p)}(\mathbf{X}) - \mathbf{X}] \, ,
%\label{dynsyss2}
\end{equation}
where $0<\lambda \ll 1$ is a scalar parameter and $\mathbf{C}$ is a constant, regular,
and real matrix.
% ps:
While the positions of the fixed points of $\mathbf{s}$ are exactly the
same as in the original system $\mathbf{f}^{(p)}$, their stability properties 
depend on the matrix $\mathbf{C}$ and the parameter $\lambda$.
The goal is to choose these parameters in such a way that the fixed point 
$\mathbf{X}_0$ to be located becomes stable with respect to the map
$\mathbf{s}$;
the more unstable $\mathbf{X}_0$ is in the original system, the smaller $\lambda$
has to be chosen in order to achieve stabilization
\cite{SchDia97PRL,DiaSchBih98PRL,SchDia98PRE,PinO00PRE}.
$\mathbf{X}_0$ is then obtained as a point of convergence when propagating a
properly chosen set of initial points with the stability transformed system
$\mathbf{s}$.
The most dominating advantages of the algorithm are the large extensions of
the basins of attraction of the individual fixed points and the fast
convergence of trajectories far from these fixed points. 

In general, a single transformation $\mathbf{s}$ is not sufficient to
stabilize all fixed points of the original system.
The above procedure is therefore applied to a set of transformations
$\mathbf{s}_i$ with different matrices $\mathbf{C}_i$ ($i=1,2,...,N$).
For our particular case of motion within the four-dimensional phase space
spanned by $(z_1,z_2,p_{z1},p_{z2})$, we find that the set of matrices
%To find POs in the four-dimensional phase space $(r_1,r_2,p_1,p_2)$,
%a set of $12$ systems $\mathbf{s_i}$ 
%sufficient for the stabilisation of all POs is
\begin{equation}
\begin{array}{cccc}
\left(\begin{array}{@{\,\!}c@{\,\!}c@{\,\!}c@{\,\!}c@{\,\!}} 
\cdot&+&\cdot&\cdot\\+&\cdot&\cdot&\cdot\\\cdot&\cdot&+&\cdot\\\cdot&\cdot&\cdot&+ \end{array}\right),
&
\left(\begin{array}{@{\,\!}c@{\,\!}c@{\,\!}c@{\,\!}c@{\,\!}} 
\cdot&\cdot&\cdot&+\\\cdot&+&\cdot&\cdot\\\cdot&\cdot&+&\cdot\\+&\cdot&\cdot&\cdot \end{array}\right),
&
\left(\begin{array}{@{\,\!}c@{\,\!}c@{\,\!}c@{\,\!}c@{\,\!}} 
+&\cdot&\cdot&\cdot\\\cdot&\cdot&+&\cdot\\\cdot&-&\cdot&\cdot\\\cdot&\cdot&\cdot&+ \end{array}\right),
&
\left(\begin{array}{@{\,\!}c@{\,\!}c@{\,\!}c@{\,\!}c@{\,\!}} 
\cdot&-&\cdot&\cdot\\\cdot&\cdot&+&\cdot\\\cdot&\cdot&\cdot&-\\-&\cdot&\cdot&\cdot \end{array}\right),
\\\\
\left(\begin{array}{@{\,\!}c@{\,\!}c@{\,\!}c@{\,\!}c@{\,\!}} 
+&\cdot&\cdot&\cdot\\\cdot&\cdot&+&\cdot\\\cdot&+&\cdot&\cdot\\\cdot&\cdot&\cdot&+ \end{array}\right),
&
\left(\begin{array}{@{\,\!}c@{\,\!}c@{\,\!}c@{\,\!}c@{\,\!}} 
\cdot&-&\cdot&\cdot\\\cdot&\cdot&+&\cdot\\+&\cdot&\cdot&\cdot\\\cdot&\cdot&\cdot&+ \end{array}\right),
&
\left(\begin{array}{@{\,\!}c@{\,\!}c@{\,\!}c@{\,\!}c@{\,\!}} 
+&\cdot&\cdot&\cdot\\\cdot&-&\cdot&\cdot\\\cdot&\cdot&-&\cdot\\\cdot&\cdot&\cdot&+ \end{array}\right),
&
\left(\begin{array}{@{\,\!}c@{\,\!}c@{\,\!}c@{\,\!}c@{\,\!}} 
+&\cdot&\cdot&\cdot\\\cdot&\cdot&+&\cdot\\\cdot&\cdot&\cdot&-\\\cdot&-&\cdot&\cdot \end{array}\right),
\\\\
\left(\begin{array}{@{\,\!}c@{\,\!}c@{\,\!}c@{\,\!}c@{\,\!}} 
\cdot&+&\cdot&\cdot\\\cdot&\cdot&+&\cdot\\-&\cdot&\cdot&\cdot\\\cdot&\cdot&\cdot&+ \end{array}\right),
&
\left(\begin{array}{@{\,\!}c@{\,\!}c@{\,\!}c@{\,\!}c@{\,\!}} 
\cdot&+&\cdot&\cdot\\\cdot&\cdot&+&\cdot\\+&\cdot&\cdot&\cdot\\\cdot&\cdot&\cdot&+ \end{array}\right),
&
\left(\begin{array}{@{\,\!}c@{\,\!}c@{\,\!}c@{\,\!}c@{\,\!}} 
-&\cdot&\cdot&\cdot\\\cdot&-&\cdot&\cdot\\\cdot&\cdot&\cdot&-\\\cdot&\cdot&+&\cdot \end{array}\right)
\end{array}
\end{equation}
with $''\cdot'' \equiv 0$, $''+'' \equiv +1$, and $''-'' \equiv -1$, is sufficient for the
identification of all fixed points.

For time-continuous systems, periodic orbits are represented by fixed points
of a suitably defined Poincar{\'e} map $\mathbf{f}(\mathbf{X})$, which permits
their detection by the ST method \cite{PinSchDia01PRE}.
In our case of a periodically driven system with two degrees of freedom,
the Poincar{\'e} map is most conveniently chosen as the stroboscopic map that
transforms the phase space variables to their propagated values after one
period $T = 2 \pi / \omega$ of the driving:
\begin{equation}
\mathbf{f}:(z_1, z_2, p_{z1}, p_{z2})|_t \longmapsto  
(z_1, z_2, p_{z1}, p_{z2})|_{t+T} \, .
\end{equation}
An initial distribution that is chosen uniform in phase space (with the
additional limitation $-5 < E < 0$ for the initial energy $E$) proved to be 
favourable for the detection of large sets of periodic orbits with
unlimited lengths. 

For polishing-up the coordinates $\mathbf{X} = (z_1, z_2, p_{z1}, p_{z2})$ of
the periodic orbits, a Newton algorithm is applied after convergence of the ST
method up to a certain accuracy $\epsilon \approx 10^{-2}$.
To this end, the monodromy matrix associated with the actual fixed point
$\mathbf{X}_0$ is approximately calculated by propagating the four points
$\mathbf{X}_i = \mathbf{X} + \eta \, \mathbf{e}_i$ ($i=1,...,4$) under the Poincar{\'e}
map $\mathbf{f}$, where $\mathbf{e}_i$ are orthogonal unit vectors in phase
space and $0 < \eta \ll 1$.
The individual columns $\mathbf{m}_i$ of the monodromy matrix
$\mathbf{M}=(\mathbf{m}_1,\mathbf{m}_2,\mathbf{m}_3,\mathbf{m}_4)$ 
are then given by $\mathbf{m}_i=({\mathbf{f}(\mathbf{X}_i)-
\mathbf{f}(\mathbf{X})})/\eta$.
This monodromy matrix is then used to improve the accuracy of the fixed point
by the Newton method.

In a similar way, the Newton method can also be employed to identify a given
periodic orbit at the field amplitude $F + \delta F$, once the corresponding orbit
at field amplitude $F$ is converged.
In that case, the fixed point $\mathbf{X}_0^{(F)}$ at $F$ would be used as
starting point for the search of the fixed point $\mathbf{X}_0^{(F+ \delta F)}$ at
$F + \delta F$
(a robust damping of the Newton algorithm by a factor $\rho \approx 0.1$ is necessary
there to ensure stable convergence \cite{ZolGre98PRE}).
This procedure turned out to be particularly efficient for identifying (within
a single run) the asymmetric stretch orbits at all field amplitudes within the
range $0 \leq F \leq 1$ (a stepsize $\delta F = 0.001$ was used in practice).
Also other periodic orbits, which were found by the ST method, were ``traced''
in this way for increasing field amplitude $F$.
While their Lyapunov exponents may exhibit local minima at finite $F$, none of
them -- except the asymmetric stretch orbit -- was found to become stable.

\bibliography{helium,PS,tunneling,local}

\begin{thebibliography}{58}
\bibitem[{lar()}]{largeB}
\bibinfo{note}{Very high magnetic field strengths $B > 2$ do still provide
  transverse confinement, but perturb the near-collinear motion so strongly
  that the configuration disintegrates along the $z$ axis.}
\end{thebibliography}

\end{document}